\documentstyle[11pt,newpasp,twoside,epsf]{article}
\begin{document}
\title{Morphological Evolution of Dwarf Galaxies in the Local Group}
\author{S. Pasetto, C. Chiosi \& G. Carraro}
\affil{Department of Astronomy, University of  Padova,
       Vicolo dell'Osservatorio 2, I-35122 Padova, Italy}

\begin{abstract}
The  dwarf galaxies of the Local Group can be separated in three
morphological groups:  irregular,   elliptical and spheroidal. 
As in the large galaxy clusters, there seems to be a 
morphology-position
relationship:  irregular galaxies are preferentially found in the 
outskirts (low density regions) of the Local Group, whereas 
dwarf ellipticals and spheroidals
are more frequent in the central, high density regions.
To cast light on the nature and origin of dwarf
galaxies in the Local Group, Mayer et al.(2001) have suggested
that a dwarf irregular galaxy tidally interacting with a galaxy 
of much larger mass may be re-shaped into a dwarf spheroidal or
elliptical object. In
this paper we check by means of N-Body Tree-SPH simulations whether
this is possible for a selected sample of galaxies of the Local Group.
Using the best data available in literature to fix the dynamical and
kinematical status of a few 
dwarf galaxies in the Local Group, we follow the
evolution of an ideal satellite, which supposedly started as an 
irregular object during its orbital motion around the Milky Way.
We find that the tidal interactions with the Milky Way 
remove a large fraction of the mass of the dwarf irregular and  
gradually reshape it  into a spherical object.
\end{abstract}

\section{Initial conditions}
We set up the initial conditions for the orbital 
motion of a dG around a galaxy of larger mass (the MW in our
case).  This task is hampered by the very small number of galaxies 
for which good determinations of  the proper motions from direct 
observations  are currently available (Irwin 1998). 
The original sample of more than 40 galaxies belonging to  the
LG drastically reduce to 5, all of them satellites of the
MW. The  orbital parameters  for this small group of objects 
are listed in Table \ref{table1}.

\begin{table}[htbp]
\centering
\caption{Components of the Galactocentric velocity vector $(V_x,V_y,V_z$), where
$V_x$ points in the direction of the sun, $V_y$ in the direction of the
rotation of the galaxy and $V_z$ to form a left-handed system. The velocities  refer
to the motion of the galaxy barycentre. The selected dwarf galaxy for our scope will be Sculptor.
 The velocities are in km s$^{-1}$.}
\begin{tabular}{l r r r}
\hline
\noalign{\smallskip}
Galaxy    &   $V_x$   &  $V_y$    &  $V_z$  \\
\noalign{\smallskip}
\hline
\noalign{\smallskip}
Sculptor   & -218.6 & +46.1   &  -243.4 \\
Ursa Minor & +27.9 & +74.1   & -227.8 \\
Draco      & +422.2 & +74.4   & -195.9 \\
Sagittarius& +204.7 & +35.6   & +281.3 \\
LMC/SMC    & +29 & -63.7  & -206.5 \\
\hline
\end{tabular}
\label{table1}
\end{table}

The dIrr galaxy, whose dynamical structure is tested against the 
gravitational interaction with a large mass object, is supposed to
have been formed in the past and to be made of three components:
DM, gas, and stars in suitable proportions. The total mass
of the galaxy is given by the sum of the three components $M_G=
M_s + M_g + M_{DM}$ with obvious meaning of the symbols.

The rotation curves (Hoffman et al. 1996) and cosmological 
arguments (Mateo 1998 and reference therein) have clearly 
indicated 
that galaxies (the dwarf ones in particular)  are DM dominated. 
Therefore we assume that $M_{DM}=0.90 M_G$. Furthermore, dIrrs are 
known to be gas-rich, with the gaseous content almost paralleling
the stellar content. We adopt here $M_g=M_s$.

The time scale required to build up the dIrr galaxy to the stage
we are considering is not relevant
here provided it is significantly shorter than the Hubble time.
Furthermore, stellar activity is supposed to be still underway
so that during the simulation more gas is converted into stars.
Finally, let us denote with $T_0$ the age of the dIrr at the start 
of the dynamical simulation. To set up  the starting model we have followed  Hernquist 
(1993), whose models have been extensively 
proved to produce 
stable configurations for systems with more than one component for a widely explained
descrition on the assumption made see Pasetto et al. (2003 in press). 

\section{Evolution of a dIrr and Metamorphosis}
\begin{figure}
\plotone{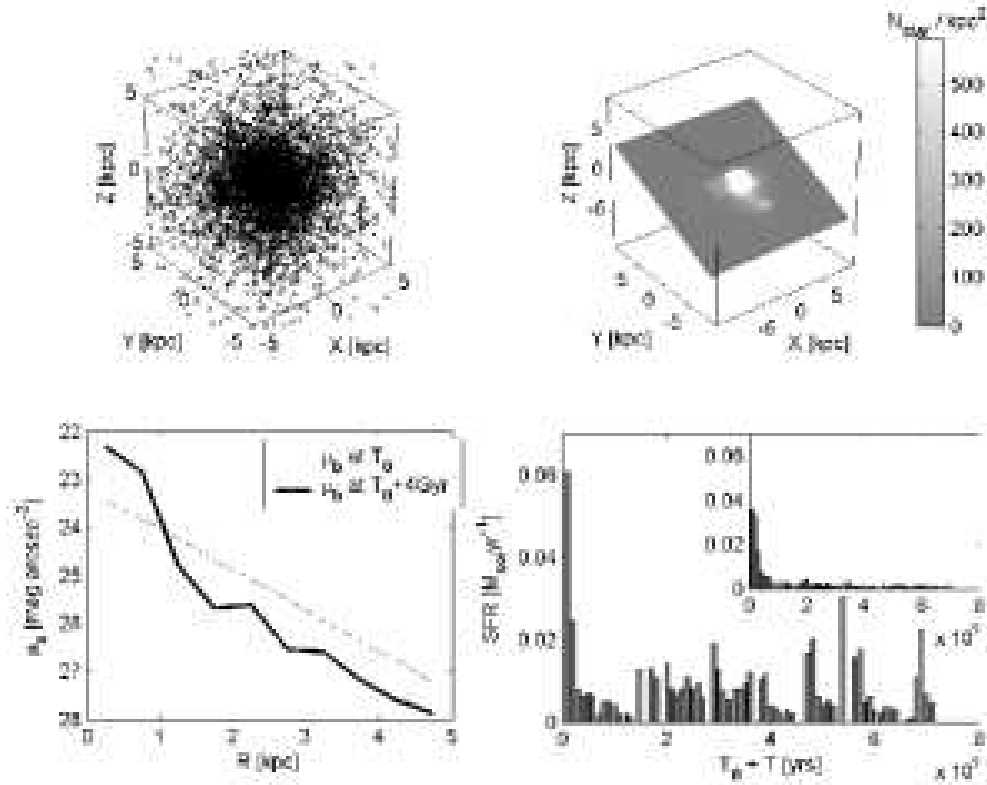}
\caption{
3D-view of the central region of the satellite galaxy after 4 Gyr of
evolution. The top left panel shows the the central body of the galaxy
enclosed in the sphere of Fig.\ref{figure1}, the tidal tail has been removed.
the remaining. The upper right panel offers a section plane with the contour of the 
iso-number density of stars in the space of position. The vertical bar displays the color code corresponding to 
different values of the number density of stars. 
The lower left panel shows the surface brigthness profile of the model galaxy at the age $T_0 + 4 Gyr$ cofronted with the initial one. The lower right panel shows the Star Formation Rate of the dG as funciton of time for the dG orbiting around and interacting with the MW. The upper sketch displays the refers to the galaxy evolved in isolation. After the initial activity, a number of small amplitude bursts of stars formation seem to occur.
}
\label{figure1}
\end{figure}

Having assigned the initial conditions, the evolution of the dIrr (disk) 
galaxy in isolation has been followed by means of the N-Body Tree-SPH
code in the last version
described by  Lia et al.(2002)
The evolution is first followed in isolation to check the dwarf galaxy model stability and then in evolution around the MW whose presence is simply reduced to a force field generated by a suitable, constant gravitational potential at the coordinate origin and with no rotation. In this study we neglect the dynamical friction as suggested in Colpi \& Pallavicini (1998) and  Colpi (1998)
As far as the  morphology of the satellite  is concerned,
two passages at the  peri-center 
are sufficient to make evident the effect we are looking for.
The situation is shown in Fig.\ref{figure1}
in which a 3-D large scale view of the satellite orbiting about 
the MW
is given at the age $T_0+4$ Gyr, i.e. after two close passages.

After the evolution we can note that

(i) A bound blob of gas, stars and DM 
with total mass of about $1.31\times 10^8 \, M_{\odot}$ and 
almost round shape survives. 
In order to better understand the morphology of the new DSph,
in Fig.\ref{figure1} we present 3D projections of  iso-number density
regions in the space of positions 
limited to the sole stellar component so that direct comparison
with isophotal contours is possible. 
 In the projection there is a distinct central concentration 
of stars with nearly spherical shape surrounded by regions of much 
lower number density. The newly born object is indeed a spheroidal galaxy.

(ii) The dSph is still DM dominated but for the very central regions 
where the baryonic mass prevails; it is almost void of gas 
(or very little) which has 
partly converted to stars and partly lost in the wind and 
tails stripped
by the MW. These properties are typical of dSph and strongly 
suggest that the kind of 
metamorphosis indicated by the simulations is at work (see Pasetto et al.).

(iii) The mayor change occurring in the dSph concerns its  kinematics.
Compared to the initial ones, all the 
three are lower by a factor of 3 indicating that the initial
circular motions have been
turned into motions equally shared along the three directions (see Pasetto et al.). 

(iv) Finally, we look at the stellar surface brightness profile (see Fig \ref{figure1}). 
Compared to the profile of the initial the surface brightness
profile of the new
dSph tends to be steeper toward the center as often observationally
indicated 
 for the dGs in the Local Group (Mateo 1998).

(v) For the case under consideration the time scale was of
the order 
of two passages past the MW (approximately 4 Gyr or shorter).
This secures {\it a posteriori} that neglecting dynamical friction, 
and consequent  evolution of the relative distance between the 
satellite 
and the MW, and finally the deformation of the MW and consequent
spiraling in of the dG towards this, 
was not too a bad approximation: see Colpi 
{\&} Pallavicini (1998), and Colpi (1998) for more 
details and a comparison with N-body simulations.

(vi) The resulting dSph seems to be stable. As a matter of fact, 
it survives  another close encounter at least at $T_0 + 6Gyrs$. Actually this helps the 
dSph to get a total mass in
closer agreement with the observational estimates for  dGs of the 
LG (Mateo 1998)

(vii) It seems therefore 
that the intensity of star formation is somehow sensitive to the 
interaction. However, as far as the correlation (if any) 
of the star forming bursts  with the orbital 
position of the satellite with respect to the MW
 (cf. Mayer et al. 2001)
it seems that the most intense ones are favored when the satellite 
galaxy is at the peri-galactic. The preliminary conclusion 
would be that   SF is mainly driven by  internal processes (see Fig.\ref{figure1}).

\end{document}